\documentclass[12pt]{article}
\title{\bf Higgs mass sum rule in the light of searching for $Z^{\prime}$ boson
at the Tevatron}
\author{
G.A. Kozlov\\
\em Bogoliubov Laboratory of Theoretical Physics,\\
\em Joint Institute for Nuclear Research,\\
\em 141980 Dubna, Moscow Region, Russia\\
\em e-mail: kozlov@thsun1.jinr.ru\\
T. Morii\\
\em Div. of Sciences for Natural Environment,\\
\em Faculty of Human Development, Kobe University,\\
\em Kobe, Japan\\
\em e-mail: morii@kobe-u.ac.jp}

\usepackage[dvips]{graphics}

\begin{document}
\date{}
\maketitle
\begin{abstract}

{\small We discuss the Higgs boson mass sum rules in the Minimal
Supersymmetric Standard Model in order to estimate the  upper limits
 on the masses of stop quarks as well as the lower bounds
on the masses of the scalar Higgs boson states.
The investigation of the bounds on the scale of quark-lepton
compositeness derived from the CDF Collaboration (Fermilab Tevatron)
data and
applied to new extra gauge bosons is taken into account. These
extra gauge bosons are considered in the framework of the
extended $SU(2)_{h}\times SU(2)_{l}$ interaction model.}

%PACS 12.38.Aw, 11.15.Kc, 12.38Aw, 12.38Lg, 12.39Mk, 12.39Pn }
\end{abstract}
%\newpage

%\section{}
%\setcounter{equation}{0}

{\bf 1}. In recent years, there are interesting discussions that
the answer to the question of why the top quark is so heavy
could be due to extra gauge interactions that single out the
fermions of the third generation.
In the simplest version of many extensions of the
Standard Model (SM), for example, the extension of $SU(2)$ gauge group
to $SU(2)\times SU(2)$ one [1-4], the massive $SU(2)$ extra gauge
bosons (corresponding to the broken generators) could couple to
fermions in different generations with different strengths.

Actually, in the model of extended weak interactions governed by
a pair of $SU(2)$ gauge groups, i.e. $SU(2)_{h}\times SU(2)_{l}$ for heavy
(third generation) and light fermions (labels $h$ and $l$ mean
heavy and light, respectively), the gauge boson eigenstates are
given as [5]
\begin{eqnarray}
\label{e1}
A^{\mu}=\sin\theta\,(\cos\phi\,W_{{3_{h}}}^{\mu}+\sin\phi\,W_{{3_{l}}}^{\mu})+
\cos\theta\,X^{\mu}
\end{eqnarray}
for a photon and
\begin{eqnarray}
\label{e2}
 Z_{1}^{\mu}=\cos\theta\,(\cos\phi\,W_{{3_{h}}}^{\mu}+\sin\phi\,W_{{3_{l}}}^{\mu})-
\sin\theta\,X^{\mu}\ ,
\end{eqnarray}
\begin{eqnarray}
\label{e3}
 Z_{2}^{\mu}=-\sin\phi\,W_{{3_{h}}}^{\mu}+\cos\phi\,W_{{3_{l}}}^{\mu}
\end{eqnarray}
for neutral gauge bosons $Z_{1}$ and $Z_{2}$, respectively which give the
neutral mass eigenstates $Z$ and $Z^{\prime}$ at the leading
order of $x$ [6]
%, respectively,
%which are defined as [6]
\begin{eqnarray}
\label{e4}
{Z\choose Z^{\prime}}\simeq
\left(\matrix{1&\frac{-\cos^3\phi\,\sin\phi}{x\,\cos\theta}\cr
\frac{\cos^3\phi\,\sin\phi}{x\,\cos\theta}&1\cr}\right)\cdot
{Z_{1}\choose Z_{2}}\ ,
\end{eqnarray}
where $\theta$ is the usual weak
mixing angle and $\phi$ is an additional mixing angle due
to the presence of two weak gauge groups $SU(2)_{h}\times
SU(2)_{l}$,
and the ratio $x$ is defined as $x=u^2/v^2$, where $u$ is the
energy scale at which the extended weak gauge group
$SU(2)_{h}\times SU(2)_{l}$
is broken to its diagonal subgroup $SU(2)_{L}$, while $v\simeq$
246 GeV is the vacuum expectation value of the (composite)
scalar field responsible for the symmetry breaking $
SU(2)_{L}\times U(1)_{Y}\rightarrow U(1)_{em}$ in the model of
extended weak interactions.
   The generator
of the $U(1)_{em}$ group is the usual electric charge operator
$Q=T_{3_{h}}+T_{3_{l}}+Y$.
% The mass eigenstates $Z_{1}$ and $Z_{2}$
%are defined as [5]
%\begin{eqnarray}
%\label{e7}
%$${Z\choose Z^{\prime}}\simeq
%\left(\matrix{1&\frac{-\cos^3\phi\,\sin\phi}{x\,\cos\theta}\cr
%\frac{\cos^3\phi\,\sin\phi}{x\,\cos\theta}&1\cr}\right)\cdot
%{Z_{1}\choose Z_{2}}\ . $$
%\end{eqnarray}
%The ratio $x$ is defined as $x=u^2/v^2$, where $u$ is the
%energy scale at which the extended weak gauge group
%$SU(2)_{h}\times SU(2)_{l}$
%is broken to its diagonal subgroup $SU(2)_{L}$, while $v\simeq$
%246 GeV is the vacuum expectation value of the (composite)
%scalar field responsible for the symmetry breaking $
%SU(2)_{L}\times U(1)_{Y}\rightarrow U(1)_{em}$ in the model of
%extended weak interactions.

At large values of $\sin\phi$, the
$Z_{2}$-boson could have an enhanced coupling to the third generation
fermions through the covariant derivative
\begin{eqnarray}
\label{e5}
D^{\mu}=\partial^{\mu}-i\frac{g}{\cos\theta}Z_{1}^{\mu}\left(T_{3_{h}}+
T_{3_{l}}-\sin^{2}\theta\cdot
Q\right)\cr
-igZ_{2}^{\mu}\left(-\frac{\sin\phi}{\cos\phi}
T_{3_{h}}+\frac{\cos\phi}{\sin\phi}T_{3_{l}}\right).
\end{eqnarray}
%To leading order, the relation (\ref{e2}) in the region where
%$\sin\phi >\cos\phi $ is
%\begin{eqnarray}
%\label{e7}
%\frac{m_{h}^2-M_{A}^2+\delta_{ZZ^{\prime}}-\Delta}{M_{Z^{\prime}}
%+M_{H}}+M_{H}- \frac{m_{W}}{\cos\phi\,\sin\phi}\,\sqrt x=0 \ ,
%\end{eqnarray}
%where $M_{Z^{\prime}}^2\simeq
%m_{W}^2\,x/(\cos\phi\,\sin\phi)^2$.

{\bf 2}. The precision measurement of electroweak parameters narrowed the
allowed region of extra gauge boson masses, keeping Higgs boson
masses to be consistent with
radiative corrections including the supersymmetric ones.
In this work, we discuss another method for estimating the
stop quark masses and upper limits on the CP-odd heavy and CP-even
light Higgs boson masses in the Minimal Supersymmetric
Standard Model (MSSM). We first show how the existing Tevatron
bounds on the scale of quark-lepton compositeness [6] can be
adopted to provide the upper limit of the quantity
$m_{{\tilde t}_{1}}\cdot m_{{\tilde t}_{2}}$, i.e. the product of
masses of stop eigenstates ${\tilde t}_{1}$ and  ${\tilde
t}_{2}$. We shall also intend to discuss how the lower bound on
the scalar Higgs bosons can be
obtained from the forthcoming Tevatron data.  It should be pointed
out that the Tevatron data [7] for
searching for the low energy effects of quark-lepton contact
interactions on dilepton production taken at
$\sqrt s=1.8$ TeV are translated into lower bounds on the masses of
extra neutral gauge bosons $Z^{\prime}$.  Furthermore, one
can emphasize that the forthcoming experiments for trying
to discover an evidence of
supersymmetry in both Higgs and quark sectors could lead us to
estimation of the masses of neutral and charged extra gauge
bosons $Z^{\prime}$ and   $W^{{\pm}^{\prime}}$, respectively.
The models in which the precision electroweak data allow these
extra gauge bosons with their masses being of the order
${\cal O}(0.5$~TeV), might be, e.g.
the non-commuting extended technicolor models [2].
The $Z^{\prime}$ and   $W^{{\pm}^{\prime}}$ bosons with such
masses are of interest, since they are within the kinematic
reach of the forthcoming Tevatron Run II experiments.

In the MSSM, the mass sum rule [8] at the tree-level
\begin{eqnarray}
\label{e6}
m_{h}^2+M_{H}^2=M_{A}^2+m_{Z}^2
\end{eqnarray}
is transformed into the following one
\begin{eqnarray}
\label{e7}
M_{Z^{\prime}}=\frac{m_{h}^2-M_{A}^2+\delta_{ZZ^{\prime}}-
\Delta}{M_{Z^{\prime}}+M_{H}}+M_{H}\ .
\end{eqnarray}
In formulae (\ref{e6}) and  (\ref{e7}), $M_{H}$ is the mass of
the CP-even heavy Higgs boson, $m_{Z}$ and $M_{Z^{\prime}}$ are the
masses of $Z$- and $Z^{\prime}$-bosons, respectively, and
$\delta_{ZZ^{\prime}}=M_{Z^{\prime}}^2-m_{Z}^2$. The correction
$\Delta$ reflects the contribution from loop diagrams involving
all the particles that couple to the Higgs bosons [9,10]
\begin{eqnarray}
\label{e8}
\Delta=\left (\frac{\sqrt
N_{c}\,g\,m_{t}^2}{4\,\pi\,m_{W}\,\sin\beta}\right
)^2\,\log\left (\frac{m_{{\tilde t}_{1}}\cdot m_{{\tilde
t}_{2}}}{m_{t}^2}\right )^2\ ,
\end{eqnarray}
where $N_{c}$ is the number of colors, $m_{t}$ and $m_{W}$ are
the masses of top quark and $W$-boson, respectively, $\tan\beta$
defines the structure of the MSSM. The values of
$\Delta\sim{\cal O}(0.01~{\rm TeV}^2)$ have been calculated [9]
for any choice of
parameters in the space of the MSSM. We suggest that the
measurement of $M_{Z^{\prime}}$ would predict
the masses of mass-eigenstates ${\tilde t}_{1}$ and
${\tilde t}_{2}$,
since $m_{t}$ and $m_{W}$ are already measured in experiment and
$m_{h}$ is restricted by the LEP 2 data [11] as $m_{h}<$
130 GeV [12]; $M_{A}$ and $M_{H}$ are free parameters bounded by
combined data coming from the MSSM parameter space and the
experimental data [13].

{\bf 3}. The Lagrangian density (LD) for an effective quark-lepton
contact interaction looks like
\begin{eqnarray}
\label{e9}
{\cal L}\supset\frac{1}{\Lambda_{LL}^2}\left [g_{0}^2
(\bar{E}_{L}\gamma_{\mu} E_{L})(\bar{Q}_{L}\gamma^{\mu}
Q_{L})+g_{1}^2 (\bar{E}_{L}\gamma_{\mu}\tau_{a} E_{L})(\bar{Q}_{L}\gamma^{\mu}
\tau_{a} Q_{L})\right ] \cr
+\frac{g_{e}^2}{\Lambda_{LR}^2}
(\bar{e}_{R}\gamma_{\mu} e_{R})(\bar{Q}_{L} \gamma^{\mu}Q_{L})
\cr
+\left [\frac{1}{\Lambda_{LR}^2}(\bar{E}_{L} \gamma_{\mu}E_{L})
+\frac{1}{\Lambda_{RR}^2}(\bar{e}_{R} \gamma_{\mu}
e_{R}) \right ]\sum_{q=u,d} g_{q}^2
(\bar{q}_{R}\gamma^{\mu}q_{R})\, ,
\end{eqnarray}
where $E_{L}=(\nu_{e},e), Q_{L}=(u,d)_{L}$; $g_{i}$ are the
effective couplings and $\Lambda_{ij}$ are the scales of new
physics. The aim of the CDF collaboration analysis [7] was to
search for the deviation of the SM prediction
in the dilepton production spectrum.
If no such deviations have been found, the
lower bound of the $\Lambda$-scale can be obtained.
The embedding of the extra gauge bosons in the model beyond
the SM gives rise
to quark-lepton contact interactions in accordance to the
following part of the LD (see [6])
\begin{eqnarray}
\label{e10}
{\cal L}\supset
-\frac{g^2}{M_{Z^{\prime}}^2}\left (\frac{\cot\phi}{2}\right )^2
\left (\sum_{l=e,\mu}
 \bar{l}_{L}\gamma_{\mu} l_{L}\right )
\left (\sum_{q=u,d,s,c} \bar{q}_{L}\gamma^{\mu} q_{L}\right ) ,
\end{eqnarray}
where $g=e/\sin\theta$.

We suppose that the couplings in the first two generations are
same in strength.
Comparing (9) and (10), one can get
the following relation between $M_{Z^{\prime}}$ and $\Lambda$ as
\begin{eqnarray}
\label{e11}
M_{Z^{\prime}}=\sqrt{\alpha_{em}}\,\Lambda\,\cot\phi/(2\,\sin\theta)\,
,
\end{eqnarray}
where the value of $\Lambda$ was constrained from the CDF
data at $\sqrt s=1.8$ TeV as $\Lambda >$ 3.7 TeV or 4.1 TeV,
depending on the contact interactions for the
left-handed electron or muon, respectively, at 95 $\%$
confidence level [6,7].

In the decoupling regime of the MSSM Higgs sector where the couplings
of the light CP-even Higgs boson $h$ in the MSSM are identical
to those of the SM Higgs bosons and thus, the
CP-even mixing angle $\alpha$ behaves  as $\tan\alpha\rightarrow
-\cot\beta$ with the $M_{A}\gg m_{Z}$ relation, one can get $M_{H}^2\simeq
M_{A}^2+m_{Z}^2\sin^{2}(2\beta)+\mu^2$ which leads to disappearance
of the $H$-Higgs boson mass in (\ref{e7}). Here,
$\mu$ is the positive massive parameter which can, in principle, be
defined from the experiment searching for separation of two
degenerate heavy Higgs bosons, $A$ and $H$. This behavior
verified at the tree-level continues to hold even when radiative
corrections are included. It has been checked that this
decoupling regime is an effective one for all values of
$\tan\beta$ and that the pattern of most of the Higgs couplings
results from this limit.

In studying the mass relation (\ref{e7}) from the extended
electroweak gauge structure, we must be aware of the issues
related to the structure of $M_{Z^{\prime}}$ in both sides of
(\ref{e7}). We suppose that $M_{Z^{\prime}}$ in the l.h.s.
of (\ref{e7}) is the mass in question to be determined
using the Tevatron data (the CDF analysis [7,6]). Therefore, one can
approximate the latter mass via the phenomenological relation (\ref{e11})
meanwhile the r.h.s. of (\ref{e7}) is model dependent where, to
leading order, the mass $M_{Z^{\prime}}$ in the extended weak
interaction model  is  $ M_{Z^{\prime}}=m_{W}\,\sqrt{x}/\cos\phi\sin\phi$
[6] in the region where $\cos\phi< \sin\phi$.
With the help of the CDF restriction for $\Lambda$ [7] entering
into (\ref{e11}), one can easily find the upper limit
on $m_{{\tilde t}_{1}}\cdot m_{{\tilde t}_{2}}$
 from the following relation
\begin{eqnarray}
\label{e12}
\Delta
<(B+M_{H}^{\star})\,(B-f\,C)+m_{h}^2-m_{Z}^2\,(1-\sin^{2}2\beta)+\mu^2\,
,
\end{eqnarray}
where $M_{H}^{\star}=(M_{A}^{2}+m_{Z}^2\,\sin^{2}2\beta
+\mu^2)^{1/2}$,
$f=f(\phi)=\cot\phi\,\sqrt{\alpha_{em}}/(2\,\sin\theta)$,
$B=B(x,\phi)=m_{W}\sqrt{x}/(\cos\phi\,\sin\phi)$, and $C$ is a
minimal value of the $\Lambda$ scale extracted from the CDF
analysis [7].
The masses of $Z$- and $W$-bosons are currently known with errors
of a few MeV each [14], whereas the mass of the top quark is known
with errors of a few GeV [14]. The dominant error on the lower
bound of $m_{{\tilde t}_{1}}\cdot m_{{\tilde t}_{2}}$
comes from the errors on mass measurements of $h$- and
$A$-Higgs bosons. In addition, the dependence of particle
couplings via $\tan\beta$ enters into the radiative correction
$\Delta$ in (\ref{e8}) and the redefinition of $M_{H}$
because of the decoupling regime. Thus,
the upper limit on $m_{{\tilde t}_{1}}\cdot m_{{\tilde t}_{2}}$
can be accurately predicted by precision measurements of the lower
bound of $M_{Z^{\prime}}$.

Fig. 1 shows the upper limit on
$L\equiv \log\left (\frac{m_{{\tilde t}_{1}}\cdot m_{{\tilde
t}_{2}}}{m_{t}^2}\right )$ for $x=2$ and 3  as a function of $\sin\phi$
at fixed values of  $\mu$ and $M_{A}$. The regions of the parameter space
lying below a
given line are allowed by the present data.
At present, the LEP bounds on the mass of
$A$-Higgs boson are $M_{A}>$ 88.4 GeV [11]. This result
corresponds to the large $\tan\beta$ region.
We see that the function $L$ is rather sensitive within the
changing of $\sin\phi$, i.e. the ratio of gauge couplings
$g/g_{l}$. Here, $g^{-2}=g_{l}^{-2}+g_{h}^{-2}$, where $g_{l}$
is associated with the $SU(2)_{l}$ group and defines the
couplings to the first and second generation fermions, whose
charges under subgroup $SU(2)_{l}$ are as in the SM, while
$g_{h}$ has the origin from the $SU(2)_{h}$ group which governs
the weak interactions for the third generation (heavy) fermions.
 In the range of $\sin\phi$ presented in the Fig. 1, the width
 $\Gamma_{Z^{\prime}}$ of the $Z^{\prime}$
falls to a minimum in the neighborhood of $\sin\phi =0.8$ [6], due
to the decreasing couplings to two first generations of
fermions. In the range $\sin\phi > 0.8$, $\Gamma_{Z^{\prime}}$
grows large, due to the rapid growth in the third generation
coupling.
%In our calculations, the parameters of
%the model are chosen as $M_{A}=0.5$ TeV (Fig.1(a)) and
%$M_{A}=0.8$ TeV (Fig.1(b)); $m_{h}=110$ GeV and $m_{h}=130$ GeV,
%$\tan\beta=30$. Here, we did not use the mass difference between
%$\tilde{t_{1}}$ and $\tilde{t_{2}}$ mass eigenstates.
%As $\sin\phi$ approaches 1, the value of
%$\log\left (\frac{m_{{\tilde t}_{1}}\cdot m_{{\tilde
%t}_{2}}}{m_{t}^2}\right )$ (see Fig.1) goes down
%to zero for small $x$.  A similar behaviour is
%seen for $x=8, 12$ and 16 when $\sin\phi$ approaches approximately

%%%%%%%%%%%%%%%%%%%%%%%%%%%%%%%%%%%%%%%%%%%%%%%%%%%%%%%%%%%

\begin{center}
\begin{figure}[h]

\resizebox{12cm}{!}{\includegraphics{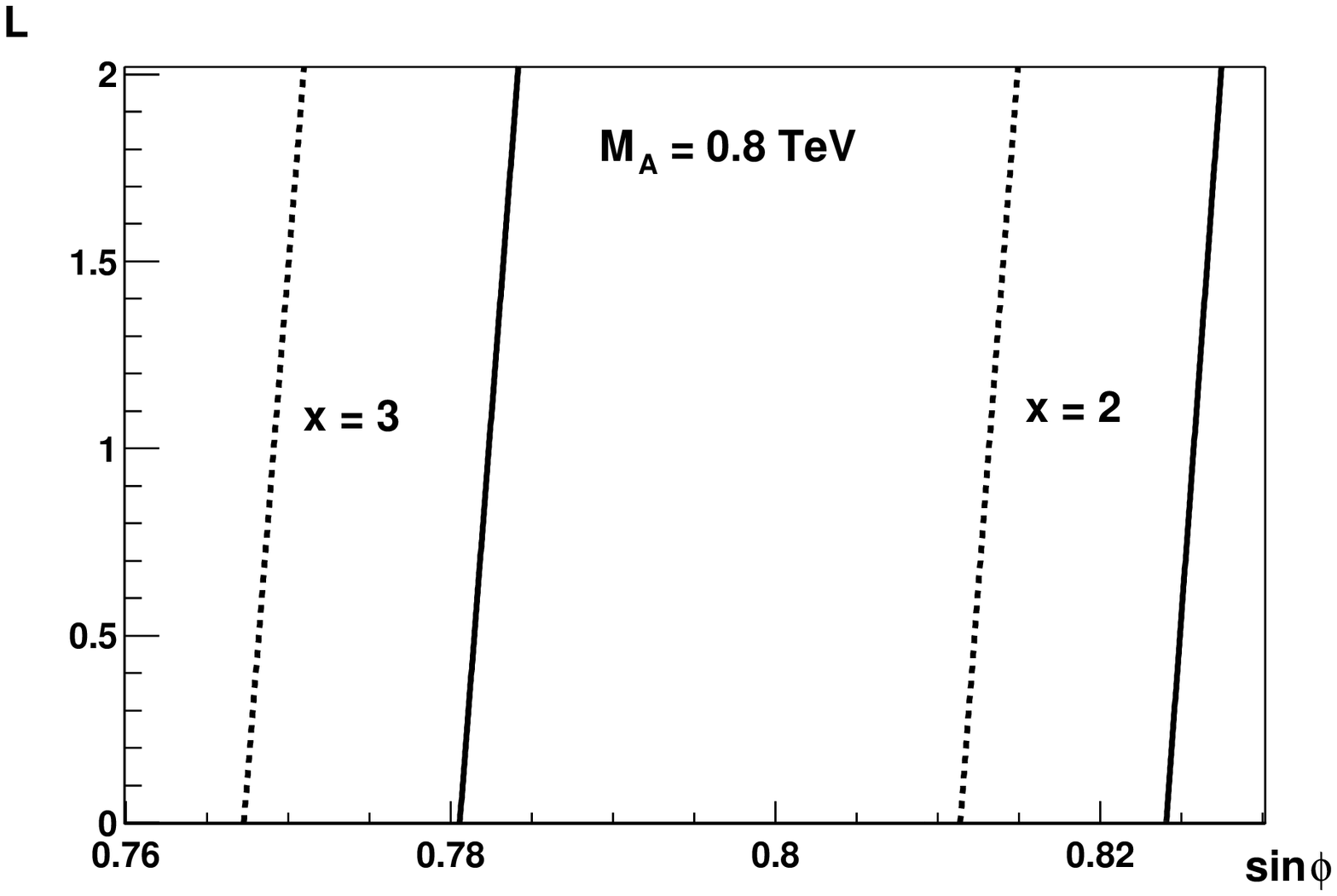}}

Fig.1 The upper limit on
$L\equiv \log\left (\frac{m_{{\tilde t}_{1}}\cdot m_{{\tilde
t}_{2}}}{m_{t}^2}\right )$
as a function of $\sin\phi$ for different values of $x = 2$ and 3;
$\mu=m_{h}=120$~GeV (dashed line), $\mu=0$ (solid line) for
$M_A=0.8$~TeV; $\tan\beta$=30.

\end{figure}
\end{center}

%%%%%%%%%%%%%%%%%%%%%%%%%%%%%%%%%%%%%%%%%%%%%%%%%%%%%%%%%%%

\begin{center}
\begin{figure}[h]

\resizebox{12cm}{!}{\includegraphics{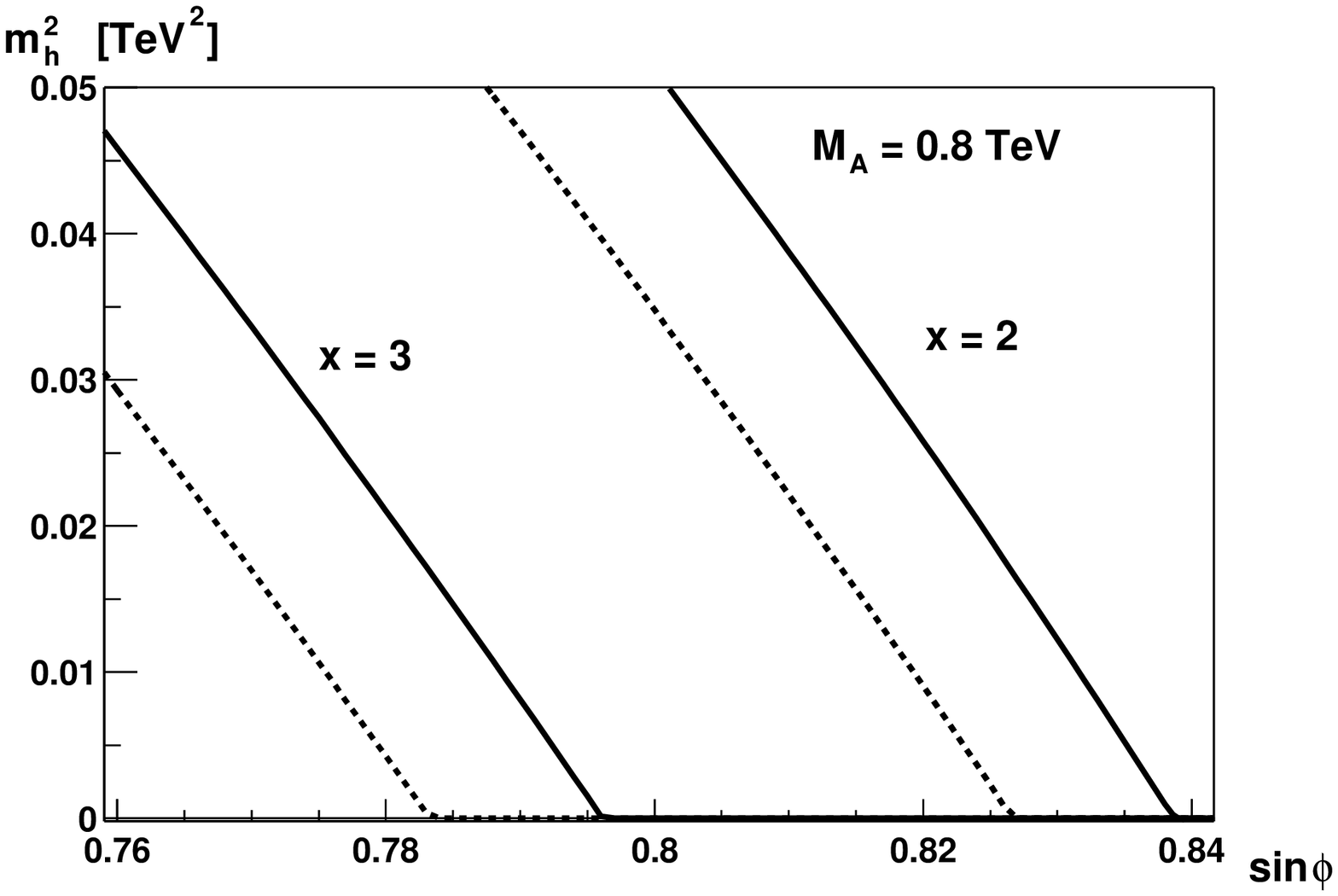}}

Fig.2 The lower bound on $m^{2}_h$ as a function of $\sin\phi$
 for different values of $x = 2$ and 3;
$\mu=m_{h}=120$~GeV (dashed line), $\mu=0$ (solid line) for
$M_A=0.8$~TeV; $\tan\beta$=30.
 The regions of the parameter space lying
above a given line are allowed by the present model.

\end{figure}
\end{center}

%%%%%%%%%%%%%%%%%%%%%%%%%%%%%%%%%%%%%%%%%%%%%%%%%%%%%%%%%%%%%%%

The CDF analysis of the contact interaction between left-handed
muons and the up-type quarks is taken into account ($C=4.1$
TeV) in our calculations.
The lower bounds of $m^{2}_{h}$ are illustrated in Fig.2.
The constraints are given for
different ratios of $x=2$ and 3 as a function of
$\sin\phi$.
In our calculations, the parameters of the model are chosen as
$M_{A}=0.8$ TeV, $m_{h}= 120$ GeV, $\tan\beta=30$.
Here, we did not use the mass difference between $\tilde{t_{1}}$
and $\tilde{t_{2}}$ mass eigenstates, and we set
$m_{\tilde{t_{1}}}=m_{\tilde{t_{2}}}=$ 1 TeV (see Fig.2).
The regions of the parameter space lying above a
given line are allowed by the present data. At the same time, a
combined fit of the experimental data [11] gives
$m_{h}=90^{+55}_{-47}$ GeV. On the other hand, recent direct
searches at LEP 2 give the
lower bound on the Higgs boson mass which is 113.4 GeV [11].

% In addition, the measurement of the
%size of $\Delta$ which is estimated to be of an order of
%${\cal O}(10^{-2}$~TeV$^2$) [9] would provide the measurement of the
%upper limits of the CP-odd $A$-Higgs boson or the lightest
%CP-even $h$-Higgs boson.
%As an example, here we estimate the upper
%limit of $M_{A}$, where other parameters are taken as
%$m_{h}=120$ GeV, $\tan\beta=30$, $\sin\phi=0.8$, and the
%ratios between two scales of the breaking symmetry are taken
%as $x=1$ and $x=2$ in the model with light $SU(2)$
%$Z^{\prime}$-boson. The
%result of this example becomes $M_{A}<$ 0.32 TeV at $x=1$ and
%$M_{A}<$ 2.3 TeV at $x=2$, respectively, for $\Delta=0.06$ TeV$^2$ and
%$C=4.1$ TeV, where $\sin\phi=0.8$ has been chosen so that
%the third generation coupling grows rapidly.

In a more extended SUSY models, their mass sum rules can give
some useful estimations with the help of the CDF data [7].
For example, in the minimal $E_{6}$ superstring theory, the
particle spectrum consists of three scalar Higgs bosons $h$,
$H_{1}$, $H_{2}$, a pseudoscalar Higgs $A$, a charged Higgs
boson pair $H^{\pm}$, and two neutral gauge bosons $Z$ and
$Z^{\prime}$. There was obtained the mass sum rule on the
tree-level in [15] in the form:
\begin{eqnarray}
\label{e13}
M_{Z^{\prime}}^{2}=m_{h}^{2}+M_{H_{1}}^{2}+M_{H_{2}}^{2}-M_{A}^{2}-m_{Z}^{2}\,.
\end{eqnarray}
The analytical expressions for the loop corrections are unknown
yet. By the way, the one-loop corrections can be summarized into
the term logarithmically dependent on the SUSY sector mass scale
[15]. Taking into account that $m_{h}$ can be identified with
the lower bound on the Higgs boson mass [11], we obtain the
lower bound on the sum $M_{H_{1}}^{2}+M_{H_{2}}^{2}$ at fixed
$M_{A}$ as a function of $\sin\phi$:
\begin{eqnarray}
\label{e14}
\sum_{j=1}^{2} M_{H_{j}}^{2}> M_{A}^{2}+m_{Z}^{2}-m_{h}^{2}+
\frac{\alpha_{em}\,C^2\,\cot^{2}\phi}{4\,\sin^{2}\theta} \,.
\end{eqnarray}

The results of the calculation of $M\equiv (\sum_{j=1}^{2}
M_{H_{j}}^{2})^{1/2}$  as the function of $\sin\phi$ is given in the Fig. 3
%%%%%%%%%%%%%%%%%%%%%%%%%%%%%%%%%%%%%%%%%%%%%%%%%%%%%%%%%%%%%%%%%%%%
\begin{center}
\begin{figure}[h]

\resizebox{12cm}{!}{\includegraphics{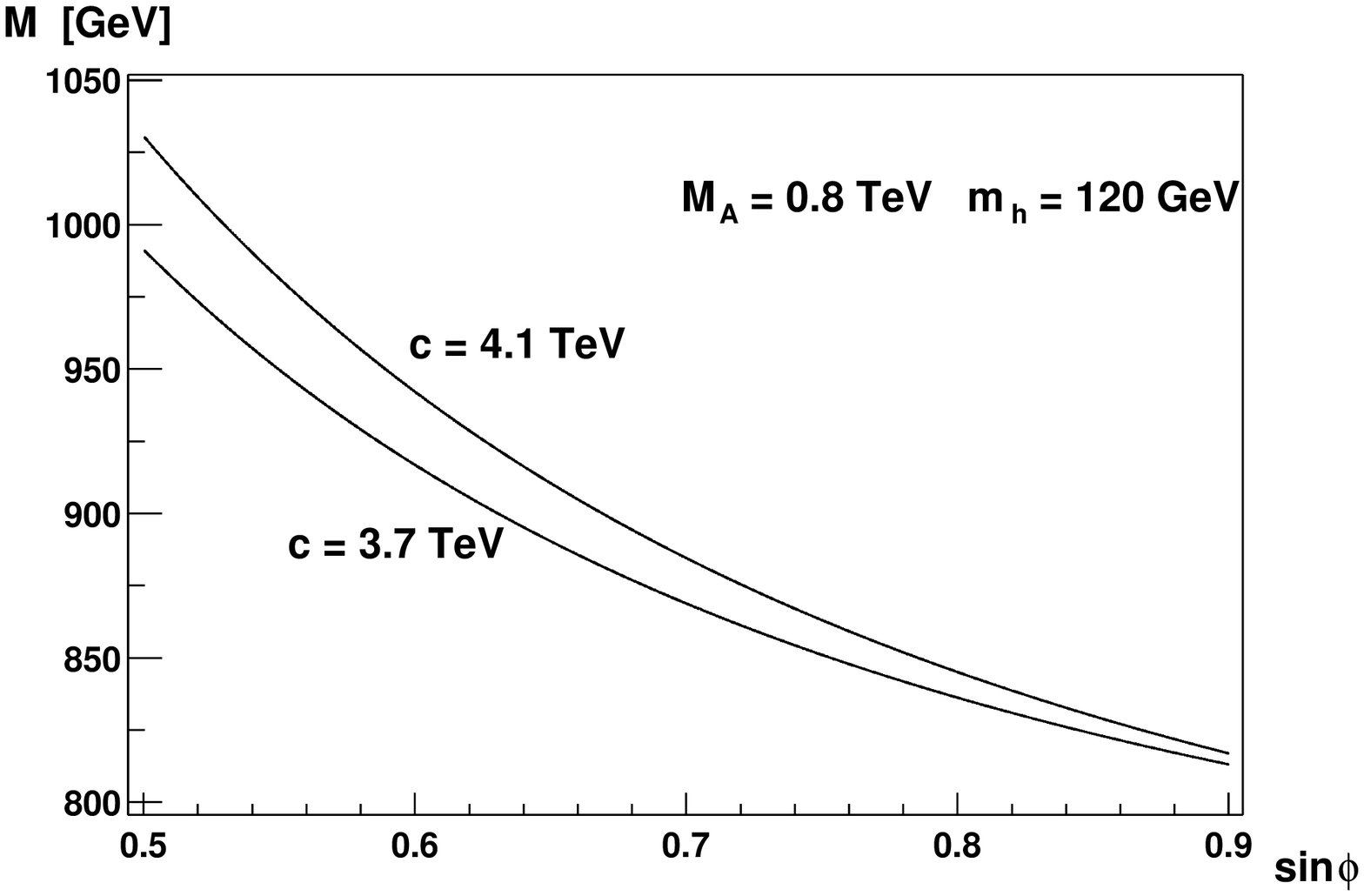}}

Fig. 3 The lower bound on $M\equiv (\sum_{j=1}^{2}
M_{H_{j}}^{2})^{1/2}$  as the function of $\sin\phi$.

\end{figure}
\end{center}
%%%%%%%%%%%%%%%%%%%%%%%%%%%%%%%%%%%%%%%%%%%%%%%%%%%%%%%%%%%%%%%

We have used the scales of new physics $\Lambda >C$ coming from the
CDF analysis [7] at 95 $\%$ confidence level: $\Lambda> 4.1 $TeV
and  $\Lambda > 3.7$ TeV for left-handed muons  and
 left-handed electrons, respectively, and up-type quarks.

{\bf 4}. To summarize, we have demonstrated that the study of
the bounds on the scale of quark-lepton compositeness derived from
the data taken at the Tevatron (CDF analysis [7,6]) and the ones
applied to $Z^{\prime}$ boson masses
within the models of the extended $SU(2)_{h}\times SU(2)_{l}$
interactions can be combined with the measurement of the upper
limits on the masses of mass-eigenstates $\tilde{t_{1}}$ and
$\tilde{t_{2}}$ and thus can sensitively probe radiative corrections
to the MSSM Higgs sector. Comparison of the experimentally
measured radiative corrections combined into $\Delta$ with its
calculations can give a precise estimation of the lower bounds
of $h$ (as well as $A$)-Higgs boson masses. The analysis of the scale
$\Lambda$  as well as the precise measurement of the lower bound
on the $Z^{\prime}$ boson mass at the Tevatron Run II can probe
the CP-violating mixing between two heavy neutral
CP-eigenstates $H$ and $A$, and as a consequence, the
non-minimality of the MSSM Higgs sector.

It is expected that the Tevatron Run II experiments will be able
to exclude $Z^{\prime}$ bosons with masses up to 750 GeV. This
leads to the restriction of the model scale parameter like $x$
which would grow. An important question is whether the
forthcoming data at the Tevatron Run II at $\sqrt s=2$ TeV will
progress far enough to determine the lower bounds on the
$\Lambda$-scale and the $Z^{\prime}$ boson mass within
the models considered in this work.\\

We thank G. Arabidze for help in numerical calculations.

\end{document}